# An Inflation Model for the Colombian Case (2001–2025)

Wilman Arturo GÓMEZ

Universidad de Antioquia[1]

Carlos Esteban POSADA

Universidad EAFIT[2]

*Abstract*

*Since the beginning of this century, the Colombian monetary authority has conducted monetary policy under a strategy based on setting targets for the central bank's interest rate and for inflation, while allowing the exchange rate of the U.S. dollar in domestic currency to float freely. This paper takes that strategy into account in order to explain inflation. Our econometric results were obtained by applying the so-called Generalized Method of Moments to test the hypotheses derived from the structural form of our model. The main findings indicate: a) the validity of a Phillips curve (that is, a positive relationship between the inflation rate and the output gap, conditional on inflation expectations); b) that the monetary authority has reacted to unexpected movements ("shocks") in inflation and in the output gap (stemming from aggregate demand surprises) by adjusting its policy in the appropriate direction but, up to the end of 2025, without being able to claim that its responses have always been timely and consistently forceful. In other words, it can be said that the monetary authority has not been aggressive in ensuring that observed inflation returns rapidly to levels consistent with the inflation target range.*

**I. Introduction**

In the wake of the 1999–2000 crisis, the monetary authority abandoned its policy of controlling the money supply and limiting fluctuations in the nominal exchange rate within a band. The replacement strategy was to allow the exchange rate to float freely (that is, to refrain from carrying out the foreign-exchange purchase and sale operations that had previously been used in an effort to impose an "appropriate" level on the exchange rate), and to adopt something that was already becoming widespread around the world: setting an inflation target and establishing an interest rate (that is, setting a rate for its short-term operations with commercial banks) at a level that would lead the actual inflation rate to converge toward a given target, while also helping the exchange rate to move, over the medium and long run, in a direction consistent with that target.

Judging by the inflation figures (Figure 1), the new strategy has been successful: between January 1991 and December 1999, the average annual inflation rate was 21.4%, whereas between January 2000 and January 2026, the corresponding figure was 5.4%. In addition, a long-run positive correlation has been observed (from January 2000 to the present) between the indices of the nominal exchange rate and consumer prices (Figure 2), indicating that the fear of exchange-rate floating, which had been common in the years prior to the implementation of the new strategy, was probably exaggerated.

---

[1] Professor, Department of Economics, Faculty of Economic Sciences, Universidad de Antioquia. Coordinator of the Applied Macroeconomics Group. E-mail: wilman.gomez@udea.edu.co
[2] Professor, School of Finance, Economics and Government, EAFIT University. E-mail: cposad25@eafit.edu.co

Figure 1

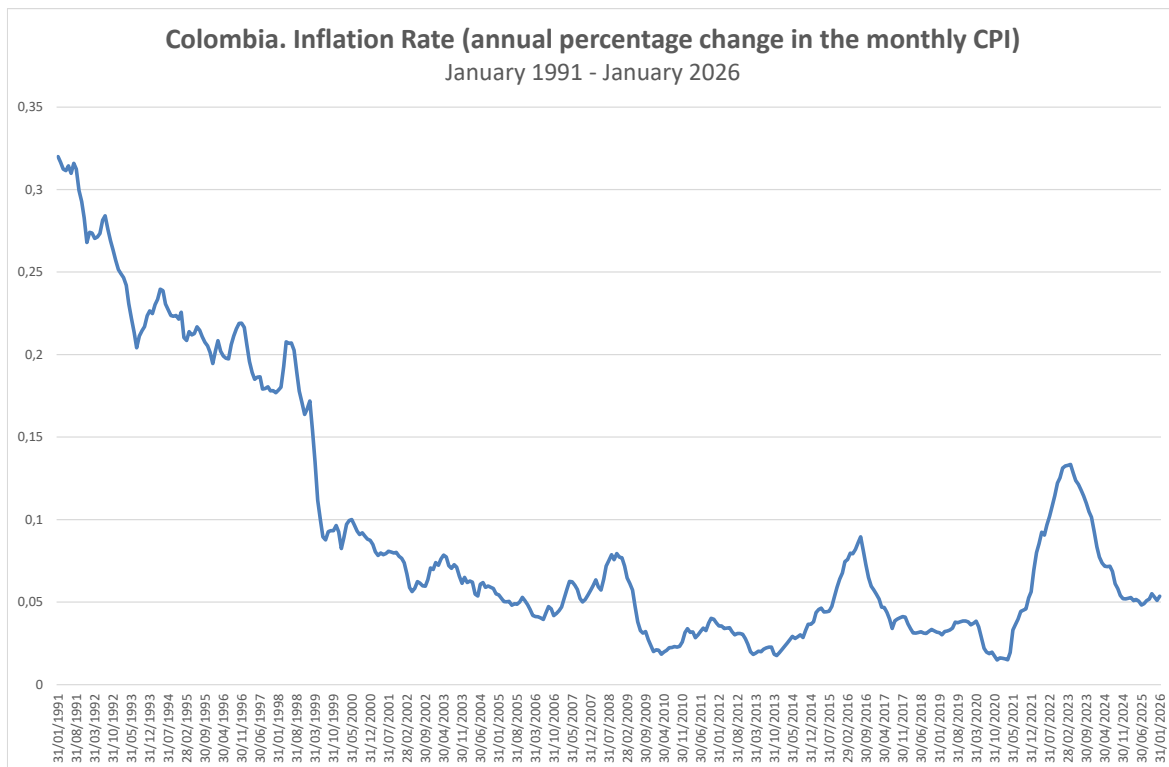

Source: **DANE** (National Administrative Department of Statistics; Colombia).

The purpose of this paper is to present a macroeconomic model of inflation and monetary policy consistent with the strategy followed in Colombia since 1999, as well as the results of an econometric estimation of the model.

In addition to this introduction, the paper is organized as follows: Literature Review (Section II); Model (Section III); Econometric Method and Results (Section IV); Conclusions (Section VI).

Figure 2

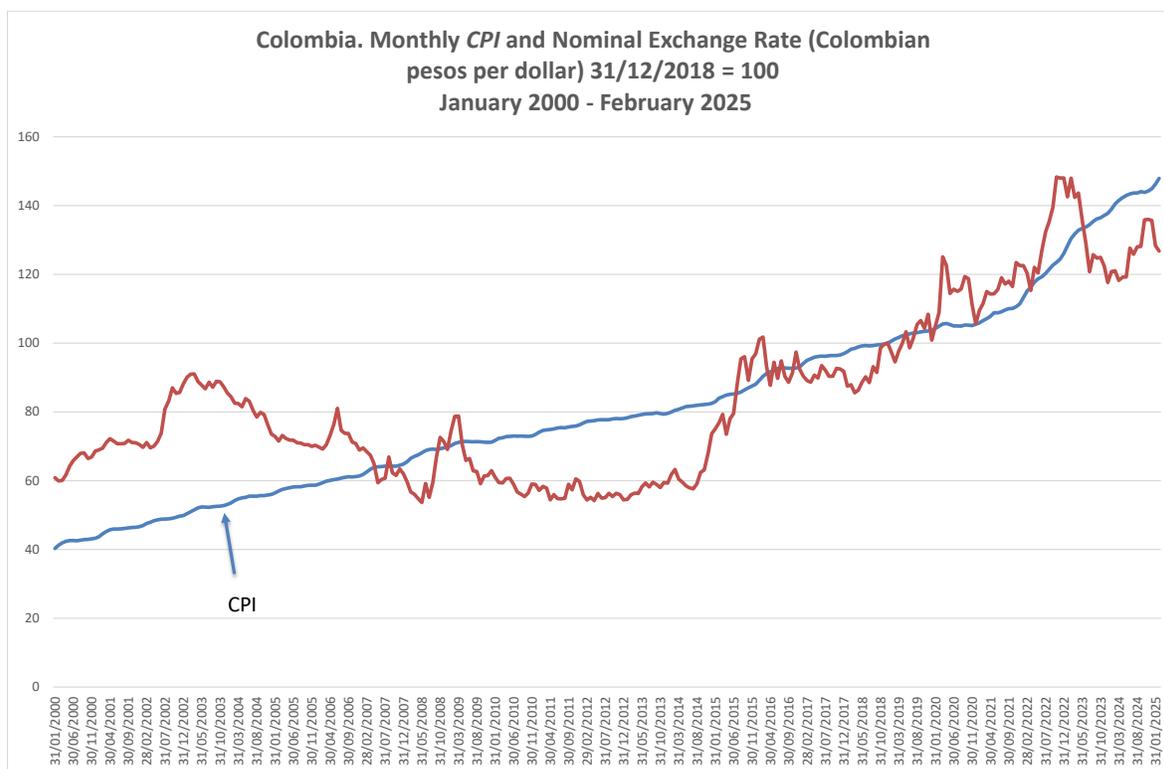

Source: **DANE** (National Administrative Department of Statistics; Colombia).

## II.   Literature

This section provides a brief overview of the literature on inflation and interest rate targeting (*I & IRT*) in small open economies, particularly in Latin America and Colombia, where the target is set by a policy interest rate.

The literature can be classified as follows: models based on forward-looking expectations (rational expectations) and backward-looking expectations (adaptive expectations) (Boug *et al*. 2015), and models in which the policy rate adheres to Taylor rules or other rules. Furthermore, *I & IRT* models assume sticky prices to theoretically justify the existence of output gaps that can persist for several quarters, and low-volatility price levels.The first contribution that should be mentioned is that of Svensson (1998): this economist was the pioneer in proposing and applying the *I & IRT* model to the case of an open economy, and he incorporated forward-looking expectations into his model; moreover, he addressed the issue of the optimality of *I & IRT* policy in the face of various shocks (random and, of course, unforeseen disturbances).

Galí and Monacelli (2005) developed a model that explicitly assumes rigid prices of the Calvo type in order to analyze how different monetary policy rules affect inflation and exchange-rate volatility.

Clarida (2014) reviewed the implications of dynamic stochastic general equilibrium models (DSGE models) for optimal monetary policy in open economies. He argued that pursuing an inflation target through Taylor-type rules can generate favorable macroeconomic outcomes, even if the exchange rate is not fully stabilized.

Finally, in the literature on inflation in small open economies, there has been debate about the relative importance of aggregate demand pressures on the inflation rate (García and Restrepo, 2001) versus that associated with external factors and the evolution of the exchange rate (López and Sepúlveda, 2022; Sussman and Wyplosz, 2024).

With regard to the Colombian case after 1999, studies on inflation have been numerous. But apparently, few of them are framed within a macroeconomic model relevant for evaluating the *I & IRT* strategy and supported by econometric results. We know of only two that may be considered predecessors of ours within this line of research, namely (in chronological order): Urrutia *et al*. (2014) and Misas *et al*. (2024).

### III. The Model

What is described in this section is an expanded and modified version of a model presented by Jones (2018) concerning the determination of the inflation rate under an *I & IRT* framework. In particular, our version, unlike Jones's, includes an expectations hypothesis appropriate for the case of a small open economy with a flexible nominal exchange rate (Galí and Monacelli, 2005; Edwards, 2006), and a (nearly conventional) Taylor rule to describe monetary policy. In its structural form, the model consists of five equations. This is set out below.

$$(1)\ \tilde{Y}_t = a_t + o_t - b(R_t - r); a_t, o_t \gtreqless 0; b > 0;$$

$$(2)\ R_t \approx i_t - \pi_t^e;$$

$$(3)\ \pi_t^e = \gamma d_t^e + (1-\gamma)\pi_{t-1}; 0 < \gamma < 1;$$

$$(4)\ i_t = r + \bar{\pi} + c_1(\pi_{t-1} - \bar{\pi}) + c_2 a_t;\ c_{i(i=1,2)} > 0;$$

$$(5)\ \pi_t = \pi_t^e + v\tilde{Y}_t + o_t; v > 0$$

Where $t$ is the time index, $\tilde{Y}_t$ is the difference between observed and potential GDP divided by potential GDP, $a_t$ is a transitory and exogenous component of aggregate demand; $o_t$ is a supply shock (transitory and exogenous); $R_t$ is the effective real interest rate (weighted average), and $r$ is the steady-state real interest rate, which we may assume to be equal to the marginal product of capital net of depreciation, $i_t$ is the nominal interest rate (weighted average; we assume it is equal to the policy rate), $\pi_t^e$ is the inflation rate expected at the beginning of period $t$, $d_t^e$ is the nominal depreciation rate of the peso expected at the beginning of $t$; and $\bar{\pi}$ is the inflation target.

Equations (1) and (2) are taken from Jones (2018, Chapters 12 and 13); equation (3) is a hypothesis about the formation of inflation expectations for the case of an open economy (a weighted average of the inflation rate in the previous period and the expected rate of devaluation); equation (4) is the expression of a monetary policy rule (which is not identical to (but is close to) the more conventional rule: the traditional Taylor rule: De Gregorio, 2007, p. 619; Romer, 2012, p. 544; and Clarida, 2014). For simplicity, the model assumes that there are only two currencies, the domestic currency and the foreign currency, and that inflation in the rest of the world is zero.

The reduced form (*RF*) is an equation that determines the inflation rate in period $t$ (the percentage change in the *CPI* over the previous 12 months) as a function of three predetermined variables and contemporaneous demand and supply shocks:

$$(FR)\ \pi_t = [(1+vb)(1-\gamma) - vbc_1]\pi_{t-1} + (1+vb)\gamma d^e$$
$$+ (c_1 - 1)vb\bar{\pi} + v(1 - bc_2)a_t + (1+v)o_t$$

The steady state is the time path along which the following conditions hold:

$$\pi_t = \pi_{t-1} = d^e = \bar{\pi};\ a_t = o_t = 0,$$

Thus, the reduced form tells us that, when the effective real interest rate converges to its steady-state level, inflation depends on four factors: the inflation rate in the previous period, expected devaluation ($d^e$), the inflation target ($\bar{\pi}$), and demand and supply shocks. In the absence of such shocks, and if monetary policymakers adhere to the policy rule (equation 4), they will bring the expected rate of devaluation into line with the inflation rate (through the interest-rate channel and through movements in the inflation rate itself), and inflation will in turn converge to the target $\bar{\pi}$. In that case, what explains the rate of increase in the price level is the inflation target. This means that, in the steady state, the only exogenous variable is the inflation target.

An approximation to the RF is the following "approximate reduced form" (ARF):

$$(ARF)\ \pi_t = [(1+vb)(1-\gamma) - vbc_1]\pi_{t-1} + (1+vb)\gamma d^e$$

$$+ (c_1 - 1)vb\bar{\pi} + \varphi\tilde{Y}_t$$

$$\text{With: } \varphi\tilde{Y}_t = v(1 - bc_2)a_t + (1 + v)o_t;$$

$$and: \varphi > 0$$

## IV. Econometric Method and Results

The *GMM* Estimator

Estimation by the Generalized Method of Moments (*GMM*) is perfect for estimating simultaneous-equation models in which there is nonlinearity, whether due to the functional form, to nonlinearity in the endogenous variables, or in the parameters. If we have a vector $f(Z;\theta)$ of n equations (linear or nonlinear) evaluated at $Z = [z_1, z_2, \ldots, z_T]$, and each $z_t$ contains the observations of a set of variables (endogenous and exogenous), the parameter vector $\theta = (\theta_1, \theta_2, \ldots \theta_q)$ contains the true, though unknown, parameters of the equations contained in $f(Z;\theta)$; thus, the unconditional expectation of this system of equations, given the parameters $\theta$, must satisfy:

$$(6) \quad E[f(Z;\theta)] = 0$$

System (6) is known as the orthogonality conditions system.

The *GMM* estimator makes it possible to choose the estimated parameters $\theta$ such that the sample analogue of equation (6), for a sample of size *T*, is as close to zero as possible:

$$\frac{1}{T}\sum_{t=1}^{T} f(Z;\theta) \approx 0$$

This system may give rise to the following cases:

1. $n = q$: the system is exactly identified, and it is sufficient to solve the system of equations (1) to (5) by nonlinear methods.
2. $n > q$: the system is overidentified (there is more information than is necessary to identify and estimate the parameter vector $\theta$).
3. $n < q$: the system is underidentified (there is insufficient information to identify and estimate the parameter vector $\theta$).

In general, cases arise in which there are more parameters than equations, which requires the use of instrumental variables or auxiliary equations in order to provide information that helps identify the parameter vector $\theta$. In that case, and given that it is important to weight the relevance of the additional information, the *GMM* estimator becomes $\hat{\theta}_{GMM}$, and it will be the parameter vector that ensures the solution to the following quadratic problem:

$$\min_{\theta} \Gamma(\theta) = F(Z;\theta)'\Omega F(Z;\theta)$$

$\Omega$ is a positive semidefinite weighting matrix that takes into account conditional heteroskedasticity of the residuals and their possible autocorrelation (DeJong and Dave, 2011; Hamilton, 1994; and Pesaran, 2015).

The empirical version of our model for *GMM* estimation is the following:

$$R_t = i_t - \pi_t^e + \varepsilon_{1t}$$

$$\pi_t^e = \gamma d^e + (1-\gamma)\pi_{t-1} + \varepsilon_{2t}$$

$$i_t = r + \bar{\pi} + c_1(\pi_{t-1} - \bar{\pi}) + c_2 a_t + \varepsilon_{3t}$$

$$\pi_t = [(1+vb)(1-\gamma) - vbc_1]\pi_{t-1} + (1+vb)\gamma d^e$$
$$+ (c_1 - 1)vb\bar{\pi} + \varphi \tilde{Y}_t + \varepsilon_{4t}$$

$$\varphi \tilde{Y}_t = v(1 - bc_2)a_t + (1+v)o_t + \varepsilon_{5t}$$

Since the version of the model to be estimated has five structural equations and 7 parameters— $b, \gamma, c_1, c_2, v, r, \varphi$ —to be identified, that is, since it is an underidentified model, this can be addressed by adding auxiliary equations. At the outset, it is natural to propose the same equations, but lagged by one period and including their products. Thus, the sample version for *GMM* estimation would consist of 23 equations (a good example of this procedure can be found in Riscaldo, 2012).

The statistical series used in our estimations were the following: the consumer price index (*CPI*) (January 2000 to February 2025); the inflation rate series is the percentage change in the *CPI* over the previous 12 months; the market exchange rate (*TRM*, by its Spanish acronym), end of month (January 2000–February 2025), as an indicator of the peso/U.S. dollar exchange rate; the series for the observed depreciation of the peso is the percentage change in the *TRM* over the previous 12 months, and the expected (monthly) rate of depreciation of the peso was assumed to be the average of the observed rates over the previous 6 months.

As for the periodicity of the variables, it should be noted that all of them, except one, are available at monthly frequency. The exception is the real GDP gap, $\tilde{Y}_t$, whose frequency is quarterly or annual. However, there is a monthly-frequency variable that may be regarded as an approximation to real GDP: the *Economic Activity Monitoring Index (ISE)*. We spliced together two series (computed by the National (Colombian) Administrative Department of Statistics, DANE: the 2005-base series and the 2015-base series), and calculated the gap $\tilde{Y}_t$ as follows:

$$\tilde{Y}_t = \frac{ISE_t - ISE\ potential\ (=trend)}{ISE\ potential}$$

The spliced *ISE* series we used covers the period from January 2000 to February 2025.

Table 1 reports the results of the econometric exercise for two different estimates of the *ISE* gap. The first estimated model uses as a proxy for the ISE gap: the difference between the logarithm of the *ISE* and its trend component, with the latter estimated using the Hodrick–Prescott filter. The second model uses the estimate of the ISE gap as the percentage deviation of the *ISE* from its trend estimated by means of a deterministic trend model.

Several relevant results may be highlighted from this exercise: (i) the results are robust to the different estimates of the *ISE* gap; (ii) all estimated coefficients are statistically significant; and, very importantly, (iii) the J-statistic test with 14 degrees of freedom (23 orthogonality conditions, combining basic and auxiliary equations, were used to estimate 7 parameters) suggests that the orthogonality conditions employed are relevant for identifying the parameters of this model.

**Table 1** *GMM Estimators*

| Parameter | Model with *ISE* Gap related to HP Filter | | Model with *ISE* Gap related to a Time Trend filter | |
|---|---|---|---|---|
| | Value | t-value | Value | t-value |
| $b$ | 0.53035 | 6.5213 | 0.53344 | 8.6989 |
| $\gamma$ | 0.063522 | 2.9909 | 0.10865 | 5.1176 |
| $r$ (long run) | 0.015771 | 30.953 | 0.016707 | 33.093 |
| $c_1$ | 1.1049 | 27.237 | 1.1356 | 23.399 |
| $c_2$ | 2.3245 | 11.878 | 2.6344 | 8.7454 |
| $v$ | 0.50492 | 7.1985 | 0.34683 | 6.4482 |
| $\varphi$ | 1.3972 | 14.607 | 1.6183 | 32.62 |
| J-statistic | 18.487 | | 22.421 | |
| 14 df $\chi^2$ critical values | | | | |
| 0.05 | 23.685 | | 23.685 | |
| 0.02 | 26.873 | | 26.873 | |
| 0.01 | 29.141 | | 29.141 | |

Source: Authors' estimates based on data from DANE and *Banco de la República*.
Note: The variables included in these estimations were found to be stationary according to the ADF test; see the Appendix

## V . Conclusions

By way of conclusions, this section summarizes and comments on the results reported in Table 1 as follows:

The value of parameter $b$ lies in the range 0.53–0.533, which, according to equation 1 of the structural model, indicates that a one-unit change in the difference between the effective real interest rate and its long-run level induces a 0.53-point change in the ISE gap in the opposite direction.

9Parameter γ (see equation 3) indicates the relative importance of expected depreciation in the expected inflation rate, which is only in the range of 6.4%–10.9%, whereas the importance of past inflation is enormous, as measured by the complementary magnitude, which lies in the range of 93.6%–89.1%.

The long-run real interest rate, $r$ (which could be regarded as a neutral rate in the absence of output and inflation gaps), is 1.6%–1.7% per year. This rate is not implausible in magnitude, but it seems relatively low (that is, we would have expected its value to lie in the range of 2%–3.5% over the last 20 years). Still, this result may be accepted unless future research, using better data and methods, compels us to reject it.

The coefficients $c_1$ and $c_2$ (see equation 4) have magnitudes that may be considered reasonable (1.1 and 2.3 or 2.6, respectively), although they are not what would be expected from a monetary authority displaying a persistent stance of strong aversion to inflation, which in colloquial terms is referred to as a hawkish authority.

In our view, the results for coefficients $c_1$ and $c_2$ are important for helping to explain why the average inflation rate over the 2000–2025 period was 5.4%, which is 35% above the upper limit of the inflation target range that the country has accepted from 2010 to the present.

Coefficient $v$ (see equation 5) indicates the sensitivity of the inflation rate to the output gap; it takes values in the range of 0.34–0.5, which should be regarded as evidence in favor of the hypothesis that a Phillips curve has existed in the Colombian case over the last 25 years.

**References**

Boug, P.; Cappelen, Å.; Swensen A. R. 2015. "Inflation Dynamics in a Small Open Economy"; *The Scandinavian Journal of Economics*; 117(3), 815–840.

Clarida, R. 2014. "Monetary Policy in Open Economies: Practical Perspectives for Pragmatic Central Bankers"; NBER WP 20545.

De Gregorio, J. 2007. *Macroeconomics*; Pearson – Prentice Hall.

DeJong, D.; Dave. C. 2001. *Structural Macroeconometrics*, (Second Edition); Princeton University Press.

Edwards, S. 2006. "The relationship between exchange rates and inflation targeting revisited"; NBER WP 12163.

Galí, J.; Monacelli, T. 2005. "Monetary Policy and Exchange Rate Volatility in a Small Open Economy"; *The Review of Economic Studies*; 72(3), 707–734.

**Anex.** Unit Root Tests

| Variable | Model specification | Coefficient for $X_{t-1}$ | P-value | ADF-statistic | P-value |
|---|---|---|---|---|---|
| GDP (ISE) gap computed from HP-Filter | Constant, no time trend | -0.266 | 0.000 | -6.830 | 0.000 |
| GDP_gap computed from regression with deterministic time trend | Constant, no time trend | -0.118 | 0.000 | -4.418 | 0.000 |
| Inflation rate | Constant, no time trend | -0.022 | 0.006 | -2.753 | 0.067 |
| Real Exchange devaluation rate | Constant, no time tren | -0.085 | 0.005 | -2.865 | 0.051 |
| *Expost Real interest rate* | Constant, no time tren | -0.018 | 0.010 | -2.595 | 0.095 |
| Nominal interest rate | Constant, no time trend | -0.015 | 0.001 | -3.420 | 0.011 |

Formal unit root tests indicate that the variables are stationary. Variables such as the *ISE* gap were estimated on the basis of the seasonally adjusted *ISE*; therefore, since they contain no seasonal component, it was not necessary to perform *HEGY* unit root tests.